\documentclass[preprints,review,accept,moreauthor,12pt]{Definitions/mdpi} 
\usepackage{aas_macros}
\firstpage{1} 
\pubvolume{8}
\issuenum{1}
\articlenumber{20}
\pubyear{2020}
\copyrightyear{2020}
\history{Received: 29 October 2019; Accepted: 18 February 2020; Published: 29
  February 2020} 
\updates{yes} 

\Title{Luminous Blue Variables }

\Author{
Kerstin Weis
~$^{1}$*\orcidA{}
and
Dominik J. Bomans
~$^{1,2,3}$\orcidB{}
}

\address{$^{1}$ \quad Astronomical Institute, Faculty for Physics and Astronomy,  Ruhr University Bochum, 44801 Bochum, Germany \\
$^{2}$ \quad Department Plasmas with Complex Interactions, Ruhr University Bochum, 44801 Bochum, Germany\\
$^{3}$ \quad Ruhr Astroparticle and Plasma Physics (RAPP) Center, 44801 Bochum, Germany
}

\corres{\hangafter=1 \hangindent=1.05em \hspace{-0.82em} {Correspondence:
    kweis@astro.rub.de}}

\abstract{ Luminous Blue Variables are massive evolved stars, here we 
introduce this
outstanding class of objects. Described are the specific characteristics, the
evolutionary state and what they are connected to other phases and types of massive
stars. 
Our current knowledge of LBVs is limited by the fact that in comparison to
other stellar classes and phases only a few ``true'' LBVs are known. 
This~results from  the lack of a unique, fast and always reliable
identification scheme for LBVs. It literally takes time to get a true
classification of a LBV. In addition the short duration of the 
LBV phase makes it even harder to catch and identify a star as LBV.
We summarize here what is known so far, give an overview of the LBV population
and the list of LBV host galaxies. LBV are clearly an important and still not 
fully understood phase in the live of (very) massive stars, especially 
due to the large and time variable mass loss during the LBV phase. 
We like to emphasize again the problem how to clearly
identify LBV and that there are more than just one type of LBVs: The giant
eruption LBVs or $\eta$\,Car analogs and the S\,Dor cycle LBVs.}

\keyword{Luminous Blue Variables; giant eruption; massive stars; stellar
  population; Wolf-Rayet stars; Eddington limit; mass loss rate; 
  nebulae of Luminous Blue Variable; Supernova impostors, bistability limit }
                    
\begin{document}


\section{Historic Background and~Naming}

Studying the brightest stars in M\,31 and M\,33 Hubble and Sandage 
\citep{1953ApJ...118..353H} found 
irregular variable stars that defined a new object class: Var\,19 in M\,31 and 
Var\,2, Var\,A, Var\,B and Vary\,C in M\,33. The~variability of Var\,2 
has been recognized already 1922 by 
Duncan \citep{1922PASP...34..290D} and 1923 by Wolf \citep{1923AN....217..475R}.
All~irregular variable stars Hubble and Sandage found showed the three common characteristics: high luminosity, blue color
indice and at the date of observation an intermediate F-type spectrum. 
Objects of this class became known
as {\it Hubble-Sandage Variables}.
In 1974 Sandage and Tammann \citep{1974ApJ...194..559S} observed bright stars 
in NGC\,2366, NGC\,4236, IC\,2574, Ho\,I, Ho\,II, and~NGC\,2403 originally 
to further constrain the Hubble constant using Cepheids. 
In some of these galaxies however they identified  stars they designated 
as  {\it Irregular Luminous Blue Variables}. 
At the same time Humphreys \citep{1975ApJ...200..426H} published 
additional spectral analysis on the
M\,31 and M\,33 Variables and put them into context to the $\eta$ Carina-like 
objects.
Few years later Humphreys and Davidson \citep{1979ApJ...232..409H} 
studied our galaxy and the LMC
and identified the most luminous and massive stars. 
In that work it became more and more obvious that a certain
region in the HRD is not populated: very luminous cool stars seems to not
exist or more likely stay for only for a very short time in this region. 
The boundary to that area was defined by the authors and has been
referred to as the Humphreys-Davidson limit. {{Shortly after in}%
~his publication entitled ``The stability limit 
of hypergiant photospheres'' de Jager~\cite{1984AA...138..246D} was the first to addressed the presence of such a
limit 
and related possible instabilities from a more theoretical perspective. 
His argumentation was based on turbulent pressure
initiating an instability. Lamers \& Fitzpatrick \citep{1988ApJ...324..279L}
however showed in a 1986 publication that ---as still accepted now---radiation and not turbulent pressure is the driver.}
This linked Humphreys and Davidson observations to the fact that
stars will become unstable in this cool and luminous~state.


The variability of S\,Doradus in the LMC was first notices by Pickering in
1897 \citep{1897PA......5..411P}, he also found the star to be bright in
H$_\beta$, H$_\gamma$ and H$_\delta$. Later further studies of its
variability \citep{1960MNRAS.121..337F}
showed that S\,Dor characteristics are very similar to those of the 
Hubble-Sandage Variables. 
{ In our own galaxy, a~class known as the P\,Cygni type stars also showed 
the same behavior. Note in that context that not all stars that show P\,Cygni line profiles
were automatically members of this historically defined class.}
Humphreys noted already in her 1975 paper~\cite{1975ApJ...200..426H} that: 
``The spectral and photometric properties of these extragalactic variables 
suggest that they may all be
related to stars like $\eta$\,Car in our Galaxy and S\,Dor in the Large 
Magellanic Cloud.''. This hints to the fact that all are only samples of one 
larger class of variable~stars.

In 1984 Peter Conti \citep{1984IAUS..105..233C} used the term Luminous 
Blue Variable during a talk at the IAU Symposium 105 on Observational Tests of the Stellar Evolution Theory.
Herewith he finally united---as Humphreys already suggested
in her 1975 paper---the earlier defined stellar subgroups of Hubble-Sandage 
Variables, 
S\,Dor Variables, P\,Cygni and $\eta$\,Car
type stars, and~explicitly excluded Wolf-Rayet stars and normal blue supergiants
from~LBVs.

\section{Characteristic of  Luminous Blue~Variables}

The name already suggests that features that LBVs seem to have in common are 
being blue and luminous stars that are variable. This however is a rather
weak constraint and not even true for a LBV all the time.
\begin{figure}[H]
\centering
\includegraphics[width=11.75 cm]{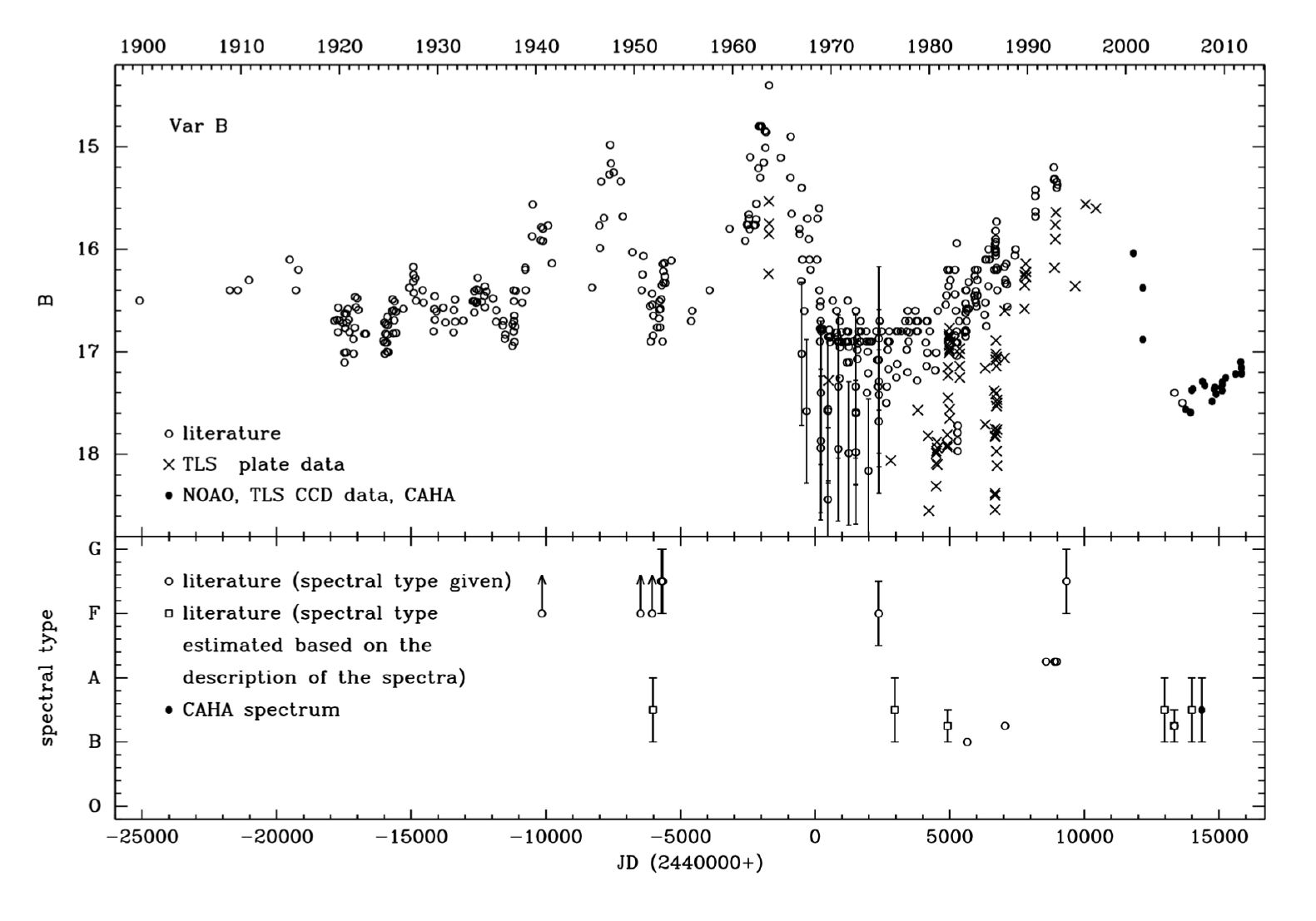}
\caption{{This figure taken from Burggraf (2015)}%
~\citep{Burggraf2015} shows a
  lightcurve  spanning more than 100 years of the LBV and original Hubble
  Sandage Variable Var B in M\,33. In~addition to the B magnitudes upper section
  the spectral type if know for the same date is plotted in the lower
  section.  Note the for S\,Dor cycles typical 
changes in the spectral type.}
\label{fig:varB}
\end{figure}

It is not simple to
disentangle a LBV from a blue O\,B supergiant and even cooler supergiant of 
spectral type A of F. A~significant number of LBVs have at least temporarily 
an Of/WN type spectrum \citep{1989PASP..101..520B, 2000PASP..112...50W}, 
indicating the presence of emission line and in particular a larger amount 
of nitrogen in their photosphere. Others were detected 
with a Be or B[e] spectrum. 
It is not possible to identify and classify an LBV by its spectrum or analog 
its color. It is the a {specific variability} or an eruption that 
distinguishes LBVs from ``normal stars''.
The variability of LBVs is a combination of a photometric brightness and 
color change, caused and accompanied by changes in the stellar spectrum. 
During such a S\,Dor variability or S\,Dor cycle which lasts years or 
decades \citep{1997AAS..124..517V, 1997AA...318...81V}
the star varies from a optically fainter to a brighter star and back. 
This variability is therefore caused by the star changing from an early (hot) 
to  a late (cool) spectral type, it implies also that not only brightens up
but also goes from a blue to a redder color. Historically the brightening of a
LBV in the bright (cool) phase during an cycle has also been called an
eruption (or S\,Dor eruption). As~we will see later this term is~confusing. 

With S\,Doradus in the Large Magellanic Cloud as the first to show this and
therefore the prototype, this alternation from hot to cool and back was 
accordingly named a S\,Dor cycle and is observed in LBVs only. \
The S\,Dor variability is the one and only
clear distinction of LBVs from  other massive evolved stars. 
An example of a long term lightcurve is given for the LBV Var\,B in M\,33 
in Figure~\ref{fig:varB}, the~analog version for Var\,C was published by Burggraf 
\citep{2015AA...581A..12B}. Also plotted here are the changes of the spectral
type for the star, that mark an S\,Dor~cycle.
\begin{figure}[H]
\centering
\includegraphics[width=9cm]{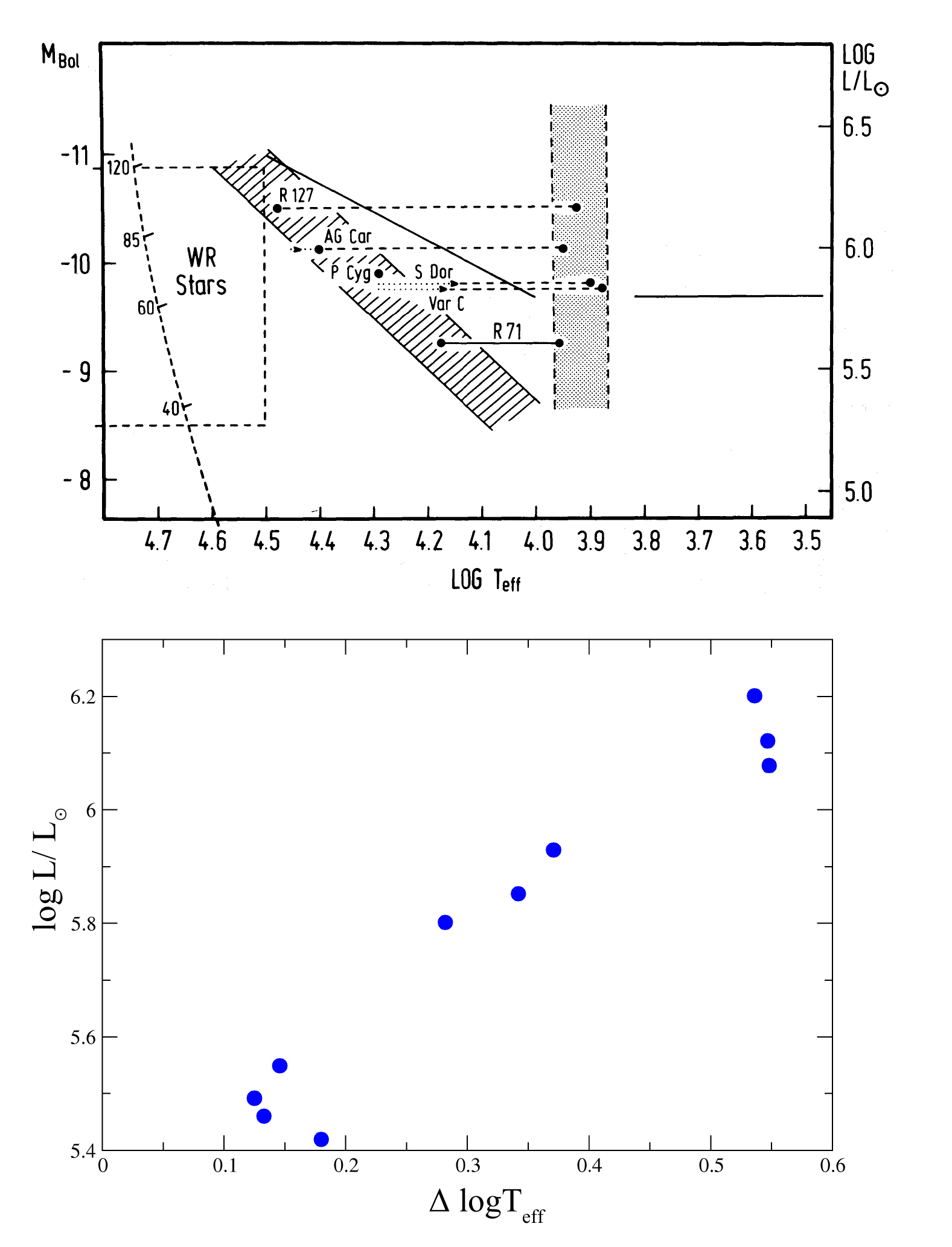}
\caption{This figure shows the classical plot by Wolf
\citep{1989AA...217...87W}  (left) and in a new version (right) we plotted the 
luminosity L and change in 
Temperature $\Delta$T$_{\rm eff}$ for a new way to visualize the amplitude-
luminosity-relation.}
\label{fig:Wolf}
\end{figure}

Bernhard Wolf \citep{1989AA...217...87W} noticed that the change of the
spectrum (or equivalent T$_{\rm eff}$) within an S\,Dor cycle from a hot 
to a cool type is larger for more
luminous LBVs. This became known as the {\it  amplitude-luminosity-relation}. 
His plot as well as a new version we made to visualize 
this relation is given in Figure~\ref{fig:Wolf}.
Instead of a classical HRD we plotted the change of Temperature
($\Delta$T$_{\rm eff}$) versus the
Luminosity L by using the LBVs given in the HRD in Figure~\ref{fig:HRD}.
The new plot visualizes nicely how tight this relation really~is.   

A more elaborate photometric classification, based on the duration of the
S\,Dor cycle was made by van Genderen in 2001
\citep{2001AA...366..508V}. He subdivided the phase and thereby objects into 
long S\,Dor (L-SD), here the cycle lasts $\leq$
20 years and the short S\,Dor (S-SD) with the cycle being less than 10 yrs. 
Beside that he added a group he designated as ex-/dormant for those that 
currently (within the last 100 years) showed only a weak or no activity at all. 
Note that these variation are much larger as the microvariability which is
common for supergiants in general~\cite{2019ApJ...878..155D}. 

In contrast to the more ordered S\,Dor variability (or eruption) LBVs 
can undergo more energetic events. In~spontaneous {\it giant eruptions} the 
visible brightness increases 
spontaneously by several magnitudes \citep{1999PASP..111.1124H}. 
The best known and well documented event is the giant eruption of the LBV
$\eta$\,Carinae around 1843. 
During the eruption the star (or rather outburst) was with $-1^{\rm m}$  the
second brightest  star in the { sky, surpassed only by Sirius} with $-1,46^{\rm m}$ 
\citep{1997ARAA..35....1D,1999PASP..111.1124H}.
Other known and documented historic and present giant eruptions of LBVs 
are those of P\,Cygni around $\sim$1600 \citep{1988IrAJ...18..163D},
SN1954J (=V12) in NGC\,2403 (\citep{1968ApJ...151..825T,1999PASP..111.1124H}, 
and SN1961V in NGC\,1058  (\citep{1964ApJ...139..514Z,1989ApJ...342..908G}). 
It is really important to distinguish between the ``S\,Dor
eruption'' and a giant eruption. The~latter being much more energetic and 
have changes of $\Delta \sim 5$mag. With~that different strength of the
``eruptions'' both are most likely caused by very different physical
mechanism. 

LBVs that showed a giant eruption 
are referred to as {\it giant eruption LBVs} or 
$\eta$ {\it\,Car Variables}, to~distinct them from LBVs that show only 
S\,Dor variations. Or~more precisely for which we at least do not know 
if they have had a giant eruption, since we are limited to historic 
records of the last centuries, several giant eruption could have 
passed unnoticed. See the contribution by  
Kris Davidson in this volume for more details on giant eruption LBVs and there
important distinction from LBVs with S\,Dor variability only. Concerning these
two very different variabilities it has so far
not been observed and therefore is not clear if the S\,Dor variability and 
the giant eruptions occur separately or a LBV can 
show both~variations.

{ Beside their variability LBVs stand out by having} a rather 
high mass loss rate. In~1997 Leitherer \citep{1997ASPC..120...58L} gave a first list for the mass loss rates of
LBVs. They range from 7 10$^{-7}$ to 6.6 10$^{-4}$ with a typical values 
around 10$^{-5}$ M$_{\odot}$ /yr$^{-1}$. Stahl~et~al. \citep{2001AA...375...54S}
used the  H$_{\alpha}$  line to determined the mass loss rate 
the during one complete S\,Dor cycle of AG\,Car. They find that the  derive 
mass-loss rates in the visual minimum is about a factor five higher as in the 
the visual maximum. More recent studies of the same object by 
Groh~et~al. \citep{2008RMxAC..33..132G} support and extend this study. The~
authors associate the changes with the bistability
limit. Lamers~et~al. \citep{1995ApJ...455..269L} first discussed  that while
evolving from hot to cool temperatures stars will pass the bistability limit at 
roughly 21000 K. At~this limit a change in the stellar wind occurs. On~the 
hot side the wind velocities are higher and the mass loss rates lower (see also 
\citep{1999AA...350..181V}). The~cool side of the bistability limit 
matches in the HRD to the region of LBVs in their cool state and causes a 
high mass loss in that phase.  
The closeness to the Eddington limit \citep{1916MNRAS..77...16E,2000AA...361..159M} of LBV in 
their cool phase also favors a high mass loss. This is even more so if 
the stars rotate fast and the modified Eddington limit the $\Omega\Gamma$
limit applies\citep{2000AA...361..159M} lowering the  gravitational force even
further. And~indeed  AG\,Car \citep{2006ApJ...638L..33G} and   
HR\,Car \citep{2009ApJ...705L..25G} are fast rotating~LBVs.

\section{The evolutionary status of~LBVs}

\begin{figure}[H]
\centering
\includegraphics[width=12.75cm]{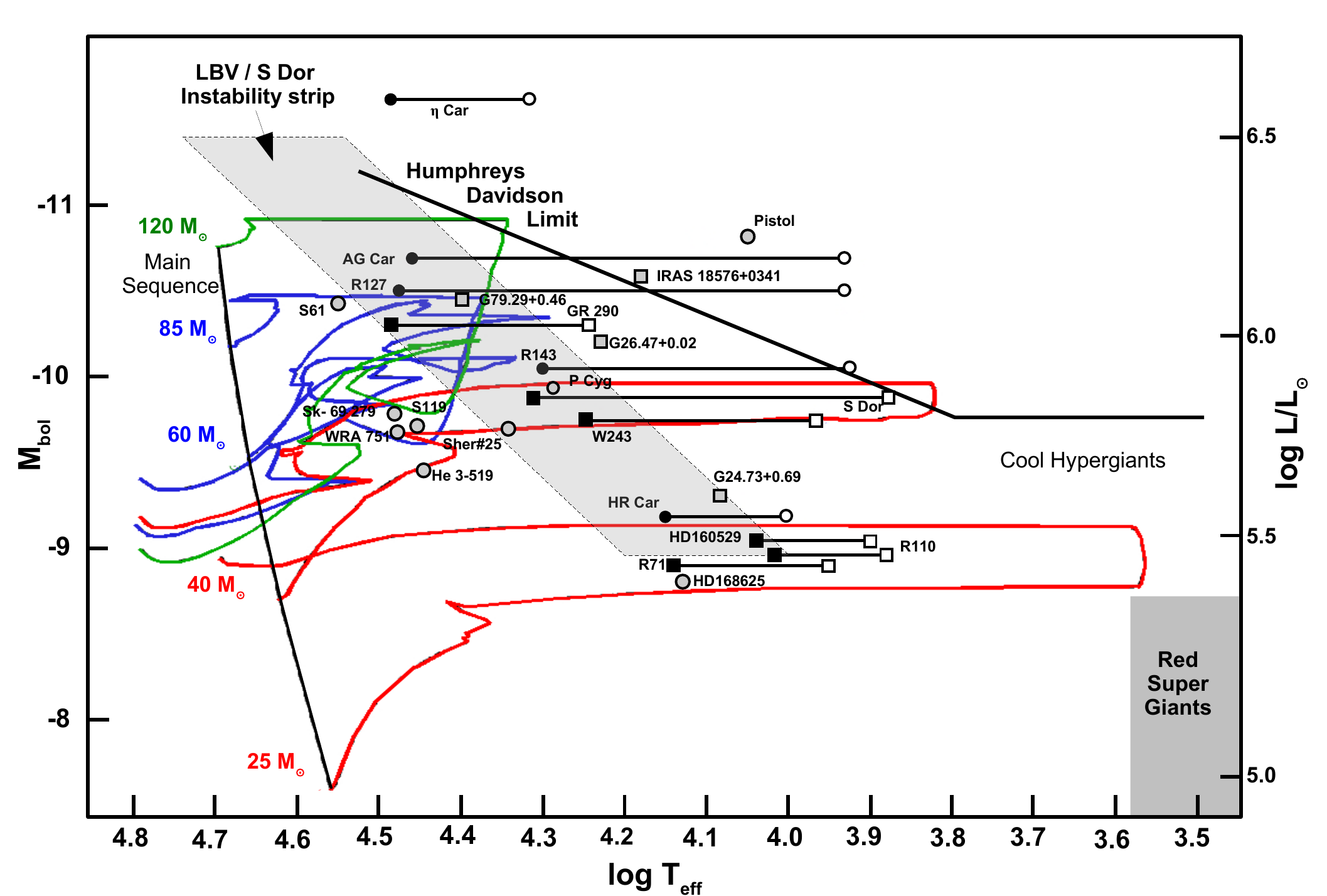}
\caption{{HRD with Galactic and LMC LBVs and LBV candidates.}
Circles are used for LBVs with an emission line (optical/NIR) nebulae, 
squares for all others. If~an S\,Dor cycle has been observed both the 
cool (open symbol) and the hot phase (filled symbol) are marked. 
Otherwise an open grayish symbol is used. In~color evolutionary tracks for 
different masses are added. The~tracks are based on the data from the Geneva
code for Z = 0.02 and v$_{rot}$ = 300 km/s, colors code the generally three
different evolutionary scenarios, see text for details.}
\label{fig:HRD}
\end{figure}

LBVs are massive evolved stars. The~LBV phase is in comparison to other
phases massive stars will pass with roughly 25000 years rather short
\citep{1991IAUS..143..485H}.  Originally, in~the classical Conti scenario 
\citep{1975MSRSL...9..193C, 1994ARAA..32..227M}, only 
stars above roughly 50\, M$_{\odot}$ were thought to turn into LBVs.
Observations however identified LBVs that have a significantly lower mass. The~
position of LBVs in the HRD, see Figure~\ref{fig:HRD} is associated with
bright and generally blue stars (like AG\,Car, R\,127, S\,61, P\,Cyg,
WRA\,751), but~an additional area is populated with LBVs that are 
fainter and somewhat cooler (HR\,Car, R\,71, HD\,160529). 
These maybe indeed hint for two subclasses the first group being massive LBVs 
and the latter less massive LBVs.
Figure~\ref{fig:HRD} shows the position of galactic and LMC LBVs and LBV 
candidates in the HRD. 
If known the position is given for both the cool (open circles) and
the hot phase (filled circles).  
Stellar evolution models by the Geneva group \citep{2005AA...429..581M} that 
include rotation also shows the position of stars with lower mass matching both 
location of LBVs in the HRD Figure~\ref{fig:HRD}.

Also plotted in this figure are tracks of the Geneva group, the~Humphreys
Davidson limit as well as the LBV/S\,Dor instability strip, the~area of LBV
in the hot~phase.

\noindent { The Geneva models \citep{2005AA...429..581M} yield} the following 
evolutionary scenarios:\\ 
\noindent least massive stars (red color code in Figure~\ref{fig:HRD}):\\
 \indent M $<$ M$_{\rm WR}$ : O -- BSG/RSG \\ 
\noindent intermediate massive stars (blue color code in Figure~\ref{fig:HRD}):\\
 \indent M$_{\rm WR}$ $<$ M $<$ M$_{\rm OWR}$ :   O -- LBV  \\
 \indent{\it or alternatively: }
\hspace{0.75cm} O -- RSG -- eWNL -- eWNE -- WC/WO \\
\noindent most massive stars (green color code in Figure~\ref{fig:HRD}):\\
 \indent M $>$ M$_{\rm OWR}$ :   O --  eWNL -- eWNE -- WC/WO \\
\noindent 
The authors define that mass limits as follows: ``M$_{\rm OWR}$ is the minimum initial mass of a single star 
entering the WR phase during the MS phase...M$_{\rm WR}$ is the minimum 
initial mass of a single star entering the WR phase at any point 
in the course of its lifetime.''
Both limits M$_{\rm WR}$ and
M$_{\rm OWR}$  depend on the rotation rate and metallicity. For~a rotation rate
of 300 km/s and solar metallicity M$_{\rm WR}$ = 22\,M$_{\odot}$
and  M$_{\rm OWR}$ = 45\,M$_{\odot}$ . Both values are higher for lower
metallicity and lower for higher metallicity.  
This leads to a mass as low as  21\,M$_{\odot}$ for LBVs at Z = 0.04. 
{ Depending on the mass and mass loss LBVs either evolve into Wolf-Rayet or
directly turn supernovae.}
Figure~\ref{fig:HRD} with the tracks and LBV positions also yield a clue to 
Wolfs amplitude-luminosity-relation: In their evolution the point in temperature
(open circle) the stars start to turned back around towards hotter
temperatures is relatively independent of the stars mass. The~more massive, luminous stars start 
with a hotter temperature, so 
for them the crossing in the HRD (or change in T$_{\rm eff}$) to the turning 
point is larger.
This cool limit is caused by the stars forming an ``extended
envelope'' or pseudo-photosphere in an opaque stellar wind
\citep{1987ApJ...317..760D, 1994PASP..106.1025H}. 
{ An~analysis of this concept using NLTE expanding atmosphere models showed 
that the formation of a pseudo-photosphere due to strongly increased mass-loss alone 
does not explain large brightness excursions \citep{Leitherer1989, deKoter1996}.  
A later discussion in context of the bi-stability jump implied that the formation 
of a pseudo-photosphere might work for rotating, relatively low mass LBVs 
\citep{Smith2004, Vink2015}. Still, the~idea of pseudo-photospheres may explain
the power-law shape of the variability spectrum at higher frequencies 
\citep{Abolmasov2011}. 
An promising alternative idea to explain S Dor variability is envelope inflation 
\citep{Grafener2012} potentially induced by changes of the stars rotation. A~similar 
idea based on an instability induced by the lowering of the effective stellar 
mass by rotation was also suggested \citep{2009ApJ...705L..25G,Groh2011}

\section{Nebulae around~LBVs}
\unskip

\subsection{Emission Line~Nebulae}

One consequence of the high mass loss rate that LBVs posses and if present 
giant eruption is the formation of circumstellar nebulae.
Many, however apparently not all LBVs are surrounded by a small nebula.
Nebulae form by wind wind interaction of faster and slower winds during a
S\,Dor cycle, while~giant eruption LBVs nebula are the result
of mass ejection in the~eruption.

LBV nebulae predominantly contain
stellar material, noticeable by the presence of stronger [N\,{\sc ii}]
emission lines as a result of CNO processed material that was mixed up into
the wind and/or ejecta of the star. During~one of the first conferences
devoted to LBVs in 1988, Stahl  \citep{1989ASSL..157..149S} reviewed on what was
known about the nebulae around LBVs.
Our current knowledge of LBV nebula is however still restricted mainly to 
nebula  in our own galaxy and the Magellanic 
Clouds, only these nebulae are are spatially resolved and can be studied in
detail. 
\begin{figure}[H]
\centering
\includegraphics[width=15.5 cm]{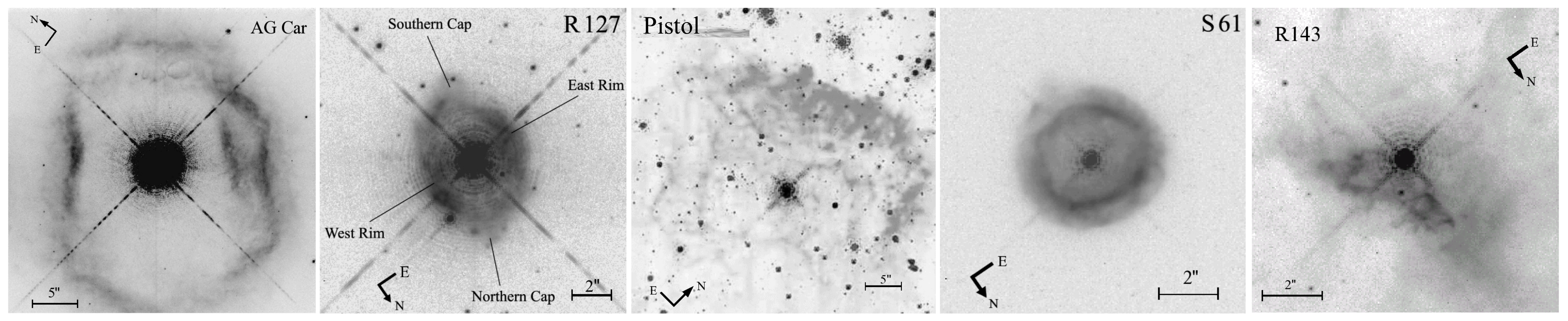}
\caption{HST images of LBV nebulae sorted by morphology: hourglass
AG\,Car \citep{2013ASPC..470..129W} , 
R\,127 with bipolar attachments, weakly bipolar He\,3-519
\citep{2015wrs..conf..167W}, spherical
S\,61 \citep{2003AA...408..205W} and last in row irregular R\,143~\citep{2003AA...408..205W}.}
\label{fig:neb}
\end{figure}   
A more recent study by Weis \citep{2011IAUS..272..372W} show that the 
morphologies of the nebula are manifold. A~
signification fraction (on average 60 \%, 75\% for galactic LBVs) 
show bipolarity. This bipolarity is either strong with a hourglass shape 
(i.e. $\eta$\,Car, HR\,Car, AG\,Car) or more weak in bipolar attachments, 
like Caps as seen in (i.e. WRA 751, R\,127). 
Figure~\ref{fig:neb} shows one example of all so far known types of morphologies 
of either a galactic or LMC nebula. 
The true bipolar nature of the nebulae around AG\,Car has been identified by
Weis \citep{2011BSRSL..80..440W}. Its hourglass structure is 
seen pol on and appears more spherical or rather boxy. Only by using high 
resolution Echelle spectra the kinematics revealed the true bipolar nature.
Only one, the~nebula around the LMC LBV R\,143 is really
irregular\citep{2003AA...408..205W}, this however is not
surprising given the stars is situated in the middle of the 30 Doradus HII
region. Spherical are S\,61 and S\,119 the latter showing signs of an 
outflow \citep{2003AA...398.1041W}.

The list with parameters in Table \ref{tab:lbvneb}
reveals that  LBV nebulae 
are with only a few parsec rather small. 
The largest is with a diameter of about 4.5\,pc the nebula 
around Sk-69$^\circ$ 279 in the LMC, the~nebula shows an 1.7\,pc extension in one
direction, enlarging the nebula size to a dimension 
of 4.5~$\times$~6.2 \,pc~\citep{2002AA...393..503W}.

The the smallest (detected so far) are the Homunculus around $\eta$\,Carinae 
(see chapter \ref{eta}), 
the inner nebula around P\,Cygni (see chapter below) and 
the nebula around HD\, 168625 (\citep{1996ApJ...473..946N}, Weis et al in prep) 
all with sizes of roughly 0.2 parsec.
Note in that context that Weis \citep{2003AA...408..205W} found for 
S\,Dor(LBV) nebula emission in the spectrum  
but it's physically to small 
to be spatially resolved (so far). The~same is true for 
GR\,290 in  M\,33 (see Maryeva this volume) and the galactic LBV W243 in
Westerlund 1. They are therefore excluded 
from Table \ref{tab:lbvneb}
and not marked bold for LBVs with nebulae in  Table Table \ref{tab:lbvlist}
The~expansion velocities of LBV nebulae are a few km/s to 100 km/s 
\citep{2005ASPC..332..271W,2011IAUS..272..372W,2012ASSL..384..171W}
They are higher for $\eta$\,Car see chapter~below. 

\begin{table}[H]
\caption[]{Parameters of Galactic and LMC LBVs and LBV candidates with 
an line emission (optical/NIR) nebulae. LBVs with dust nebulae only have been 
excluded here. In case the nebula has several spatially distinct parts (inner and outer regions) a slash is used for separation between them.}
\label{tab:lbvneb}
\begin{center}
\begin{tabular}{|l|l|l|l|l|l|}
\hline 
\textbf{LBV} & \textbf{Host Galaxy} & \textbf{Maximum Size} & \boldmath{$v_{\rm exp}$}  & \textbf{Morphology} & \textbf{Reference} \\ 
 &  & \textbf{[pc]}  & \textbf{[km/s]} & &    \\ 
\hline
$\eta$\,Carinae &  Milky Way  & 0.2/0.67  & $300$/$10-3200$ & bipolar &
\citep{2001ASPC..242..129W,2005ASPC..332..271W} \\
AG\,Carinae & Milky Way & 1.4\,$\times$\,2 & ${25}/{43}$ &
bipolar & \citep{2008ASPC..388..231W}\\
HD\,\,168625 & Milky Way & 0.13\,$\times$\,0.17 & 40 & bipolar &
\citep{1996ApJ...473..946N}\\ 
He\,3-519 & Milky Way & 2.1 & 61 & spherical & \citep{2015wrs..conf..167W} \\ 
HR\,Carinae & Milky Way & 1.3\,$\times$\,0.65 &  $75$ & bipolar & \citep{1997AA...320..568W} \\
P\,Cygni &  Milky Way & 0.2/0.8 & $110-150$/185 & spherical & see text \\
WRA\,751 & Milky Way & 0.5 & 26 & bipolar & \citep{2000AA...357..938W} \\ 
Pistol star & Milky Way &  0.8\,$\times$\,1.2 & 60 & spherical & \citep{1999ApJ...525..759F}\\ 
Sher\,25 &  Milky Way & 0.4$\times$1 & $20/83$ & bipolar &
\citep{1997ApJ...475L..45B}  \\ 
R\,127 & LMC & 1.3 & 32 & bipolar & \citep{2003AA...408..205W} \\ 
R\,143 & LMC & 1.2 & 24 (line split) & irregular & \citep{2003AA...408..205W}\\ 
S\,61 & LMC & 0.82 & 27 & spherical & \citep{2003AA...408..205W} \\ 
S\,119 & LMC & 1.8 & 26 & spherical plus outflow & \citep{2003AA...398.1041W} \\ 
Sk-69$^\circ$ 279 & LMC & 4.5$\times$6.2 & 14 & spherical plus outflow & \citep{2002AA...393..503W} \\
\hline
\end{tabular}
\end{center}
\end{table}

\subsection{Dust~Nebulae}

With the SPITZER MIPSGal survey more than 400 small ($\sim 1'$) single
bubbles were detected  in $24\,\mu$m emission~\cite{Mizuno2010, Simpson2012}.
An extended sample was even derived using citizen-science and machine-learning
methods~\cite{Beaumont2014}. 
For most of these bubbles  no optical counterpart is known, making
them heavy obscured gas and/or pure dust bubbles.   Some of the bubbles
contain central, NIR bright stars (some even faintly visible in the optical),
while others do not show central sources at all, not~even in SPITZER IRAC
images $3.6\,\&\,4.5 \mu$m  from the GLIMPSE surveys.  The~nature of these small
bubbles were an enigma, until~first classification spectra of some  the bright
central sources were taken~\cite{Gvaramadze2010, Wachter2010}.
Several of those turned out to be massive evolved stars, like
blue supergiants, LBV candidates and Wolf-Rayet
stars. Others were red supergiants and AGB stars.  The~nature and origin of
the emission of the bubbles however remained uncertain. It could be hot dust,
or MIR lines of ionized gas, or~both.  Taking SPITZER IRS spectra of several
of the bubble revealed that all cases exist~\cite{Flagey2011, Nowak2014}:
bubbles for which no central stars are detected seems to be
dominated by line  emission (mostly the high ionization [OIV] $\lambda
25.9\mu$m line), and~are therefore most likely  planetary nebulae.  
Bubbles with NIR visible central stars that show dust dominated IRS
spectra are even less frequent. 
Stellar NIR spectroscopic classification again prove
that the central stars are dominated by evolved massive stars~\cite{Flagey2014} of various types, with~several Wolf-Rayet stars,  e.g.~\cite{Gvaramadze2010_2, Mauerhan2010}, and~ a number of LBV candidates or
related stars~\cite{Gvaramadze2012, Gvaramadze2014, Gvaramadze2015}.
Two of these candidates can now be seen as established LBVs (see Table~\ref{tab:lbvlist}, 
WS1 \citep{Kniazev2015} and MN48 \citep{Kniazev2016}.   Also, several previously known 
LBVs show MIR nebulae, e.g.\ MWC 930 \citep{Cerrigone2014} or HR\,Car \citep{Umana2009}.

Still there are 
some problems with interpreting the small MIR bubbles
and especially the LBV candidate interpretation.
Most of these bubbles are
round/spherical, e.g.~\cite{Gvaramadze2010}, and~bipolar structures are
rare among the $24\mu$m bubbles.  While this is consistent with the
morphology of circumstellar gas nebulae of Wolf-Rayet stars, it seems to contradict 
the results found for LBV nebulae~\cite{2003AA...408..205W} which have as
reported above a preference for bipolar morphologies.
From the $24\,\mu$m images
alone it is not clear, whether the nebulae are a) dust only b) partly dust,
partly gas, or~c) dominated by ionized gas.  The~distribution and kinematics
of the dust and gas are often different for circumstellar  nebulae of massive
stars, as~e.g.,~shown for the nebula of the classical LBV AG\,Carinae~\cite{2008ASPC..388..231W}. 
{ Gail~et~al.~\cite{2005ASPC..332..317G} were among the first to 
investiage the problem of dust formation in an CNO precessed material 
like LBV envelops. Later hydrodynamical simulations of gas and dust nebulae,}
e.g.\ \cite{vanMarle2011}, show the small dust grains follow the gas quite well,
but the larger grains show their own unique distribution.
The morphologies problem may therefore be dominantly a wavelength bias.
Last but not least one can speculate about the bubbles being signposts of a more
general, previously overlooked short high mass-loss phase in the evolution
of many massive stars. 
We are currently running a program with the LBT infrared spectrograph LUCI to
classify more of the central stars using HK band spectroscopy.

\subsection{P\,Cygni}

As was mentioned above P\,Cygni is one of the classical giant eruption LBVs, 
with an eruption observed in 1600.In the Van Genderen classification it is 
currently a weakly active LBV.

\begin{figure}[H]
\centering
\includegraphics[width=10 cm]{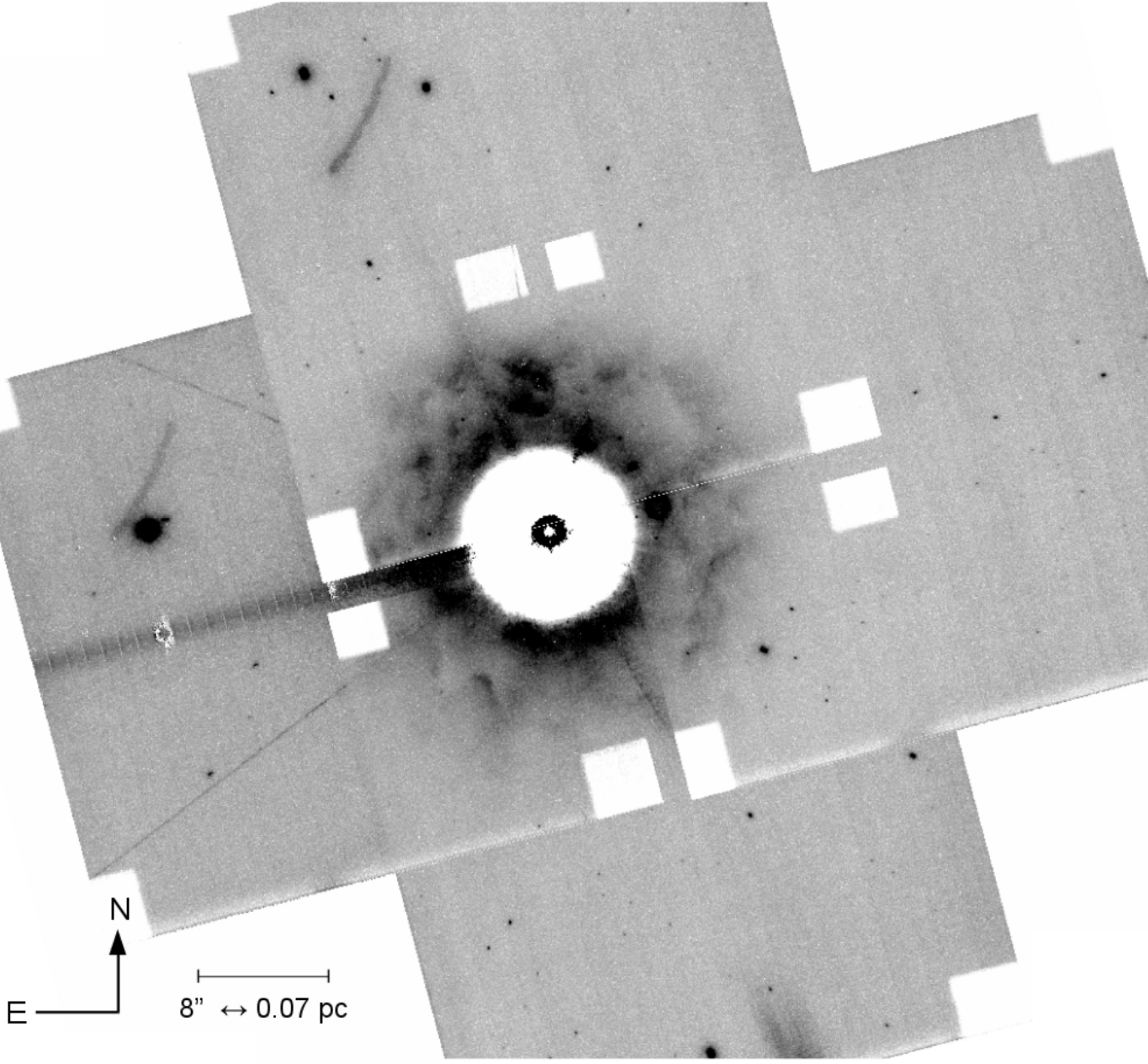}
\caption{An LBT LUCI AO [FeII] image of the inner nebula (or inner shell)
of P\,Cygni (Weis~et~al., in~prep). The~images has pixel scale of 0.015''/pixel 
and resolve scales down to ~ 85 AU.} 
\label{fig:Pcygni}
\end{figure}

The nebula around P\,Cygni has at least,
two distinct parts. One larger structure with a 
diameter of 0.8 to 0.9 pc named the outer shell or OS which is rather 
spherical and a much smaller and clumpy structure the inner shell or IS 
which is less than 0.2 pc across \citep{1994MNRAS.268L..29B}.
Beside the IS and OS
Meaburn~et~al. \citep{1999ApJ...516L..29M} reported 1999 a giant lobe to
associated with the stars. Its has a PA = 50$^\circ$ to the other nebulae and
stretches to an extend of 7 arcminutes or up 3.6 pc. He finds expansion 
velocities around 110 to 140 km/s (depending on which line he uses) associate 
with the inner shell and structures as high 185 km/s in the outer shell.
With a diameter of only 0.2\, pc the IS is in most
images barely resolved. A~new LBT/LUCI AO image (Figure \ref{fig:Pcygni}) we
made recently shows the large amount of fine structure and details of the inner nebula for the
very first time. With~a resolution down to 85 AU size structures 
it is an improvement from the previously published LBT image by Arcidiacono~et~al. \citep{2014MNRAS.443.1142A}.

Mapping the nebulae with KPNO high resolution longslit Echelle Spectra we 
measured  the expansion velocity  of the inner nebula is 100--150 km/s, this
is well in agreement to  Meaburns values. This would assuming no larger
acceleration or deceleration match to the inner  shell having been  ejected
during the 1600 giant eruption. With~the spectra we can also associate
velocities to distinct  clumps that appear in the spectra and can be 
identified on the image (Weis~et~al. in prep.).

\subsection{$\eta$\,Carinae --- the Most Peculiar LBV?}\label{eta}

$\eta$\,Carinae used to be the most classical giant eruption LBV 
or $\eta$\,Carinae variable. With~the discovery of many unique and unusual
characteristics  $\eta$\,Carinae or the $\eta$\,Carinae system is not the 
LBV par excellence anymore. A~book devoted to $\eta$\,Carina and the 
Supernova impostors \citep{2012ASSL..384.....D} can be consulted for all
details on this object. Here a short summary of the characteristics:
\begin{itemize}[leftmargin=*,labelsep=5.5mm]
\item[$\bullet$] A giant eruption that took place in~1843 
\item[$\bullet$] A binary system with two massive 
components one with 60 M$_{\odot}$ the second with 30 M$_{\odot}$ 
\citep{2012MNRAS.420.2064M}  
\item[$\bullet$] A nebula that has at least three section: 
{\it The little Homunculus}, {\it The Homunculus},
{\it The outer ejecta}.
\end{itemize} 


\begin{figure}[H]
\centering
\includegraphics[width=14cm]{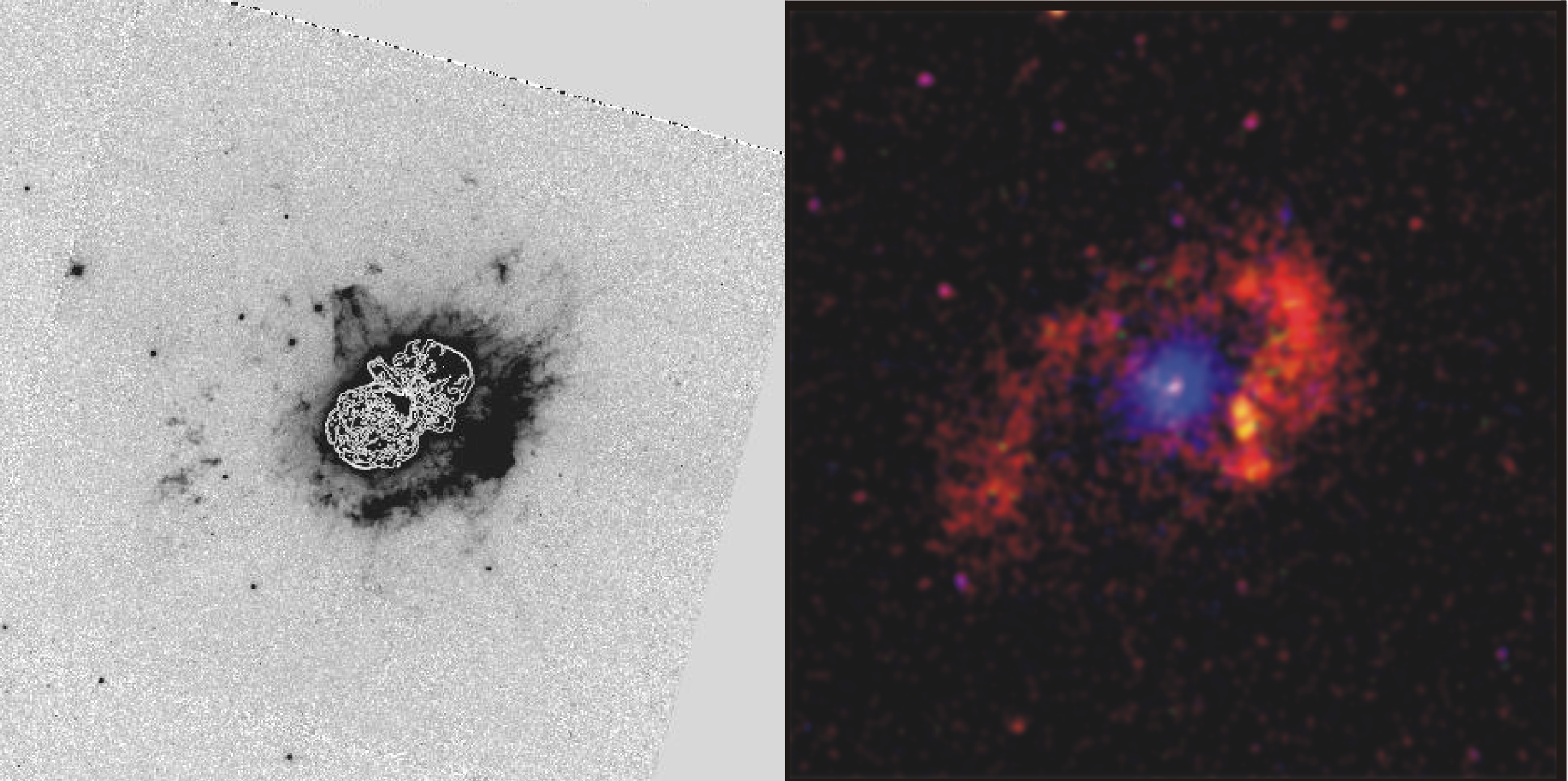}
\caption{The nebula around $\eta$\,Carinae in the optical and X-ray. 
Left: An optical F658N HST image in greyscale, the~Homunculus nebula 
additionally marked in contour to distinguish it from the outer ejecta, shown 
only in grey scale \citep{2005ASPC..332..271W}
Right: A CHANDRA Xray image with color coded energy regimes, green:0.2-0.6
keV, red: 0.6--1.2 keV and blue 1.2--12 keV color version of Figure~1 in 
\citep{2002ASPC..262..275W}.
}
\label{fig:eta}
\end{figure}

The Homunculus was identified first and photographed in 1950 by 
Gaviola \citep{1950ApJ...111..408G}, the~name of the nebula was motivated 
by the first images showing a man like morphology.   
The little Homunculus resides within the Homunculus and was revealed only using 
HST STIS long slit spectroscopy by Ishibashi~et~al. \citep{2003AJ....125.3222I}. The~outer ejecta as the name implies
surrounds the Homunculus. It consists of a countless 
number of clumps and filaments. A~first report and catalog with designation of
several part of the outer ejecta was made 1976 by 
Walborn \citep{1976ApJ...204L..17W}.A~summary of more recent optical, x-ray
and kinematic  studies of the outer ejecta is given by 
Weis \citep{2012ASSL..384..171W}.

Today we know that all three sections of the nebula are of bipolar morphology.
The expansion velocities are with up to 3000 km/s 
faster than in any other LBV nebula. Shocks of these extremely fast 
structures in the 
outer ejecta create  X-ray emission \citep{2001AA...367..566W,2004AA...415..595W}. 
This emission is shown in the right section of Figure~\ref{fig:eta}, here a 
a CHANDRA image is color coded and in indicates in red soft Xray emission
of the outer ejecta, in~blue the more central emission results from 
shocks of the central stellar system {not} the Homunculus nebula!

\section{Instabilities and the Origin of~Variability}

What are possible origins of the LBV variability. First we have to
differentiate between the S\,Dor variability and giant eruptions. The~latter
are in need of much larger energy being released. 
Already~in~their 1994 paper Humphreys and Davidson \citep{1994PASP..106.1025H} 
discussed  what could cause the variabilities and whether one or more mechanism
are at work. They argue and cite several works showing that a classical
$\kappa$-mechanism seems not to work. A~more likely cause also discussed in that
paper is the proximity of LBV to the Eddington, or~in case of rotation
$\Omega\Gamma$ limit.
This limit indeed lies in the HRD in the same region as the Humphreys Davison
limit, which also resembles the cool position of an LBV in the S\,Dor cycle. 
{Clearly the properties of the stellar winds and their dependence of metallicity 
have to be a major contributor to the mechanisms creating the variability.}
For various more detailed theoretical works the reader is referred a review by
Glatzel \citep{2005ASPC..332...22G}  a newer overviews by Vink
\citep{2012ASSL..384..221V} and Owocki \citep{2015ASSL..412..113O}. 
Alternative models like non-radial gravity mode
oscillations have been proposed by Guzik \citep{2005ASPC..332..204G}. 
The potential importance of pulsations for the driving of the S\,Dor mechanism
was discussed  in \citep{Lovekin2014}.  The~analysis of long, well sampled
lightcurves may provide  more information about the properties of the S\,Dor
process  \citep{Kalari2018, 2019ApJ...878..155D}.  Still, the~link of low
amplitude variability  patterns with LBV nature is far from
clear. Kalari~et~al.\ \citep{Kalari2018}  showed that SMC blue supergiant
AzV\,261 exhibits variability patterns consistent with  other LBVs in a 3 year
time span and nearly nightly photometric observations, but~their  detailed
analysis of high dispersion spectra showed no temperature changes typical for
an S\,Dor cycle over a decade, precluding a classification as~LBV.

Recently, new 3d radiation hydrodynamic simulations of 80 and 35\,M$_{\odot}$
performed and the results point at variations of the He opacity as a possible cause of
the  S Dor variability and link the shorter time scale irregular oscillations
to convection  \citep{Jiang2018}. Other ideas for the origin of the S Dor
variability were already discussed  in chapter 3.

\section{The LBV Wolf-Rayet Star~Connection}

In the last years it has been found that the masses of Wolf-Rayet stars in that
state (not their initial mass) is much lower as can be explained by the stellar
winds only. Furthermore the { empirical} mass loss rates for hot, massive stars 
have also been seriously questioned, mainly because of the effects of wind clumping
\citep{2008cihw.conf.....H,2011AA...528A..64S}. Wind clumping will reduce the mass 
loss rate and leave us with even higher mass in evolved stars.  { Stellar evolution 
models use theoretical mass loss rates, generally lower that the empirical ones even 
without clumping.}  A phase of enhanced mass loss { with a different mechanism 
 may be needed \citep{Puls2008}.}  

The LBVs phase would just fit. It is passed right before the WR phase 
and is known for high mass
loss as well as the formation of massive nebulae. A~LBV phase therefore might
be mandatory to explain at least some WR classes and the lower WR star masses. 
One might even speculate that WR nebula are only by fast WR winds blown up,
enlarged former LBV nebulae. Indications for such a hypothesis are the somewhat
larger size of  WR nebulae in combination with a N enhancement. 
The~latter being a well known
attribute of LBV nebula. Most WR nebula are not found around WO or WC  but 
WN type stars, the~natural and direct predecessor of~LBVs. 

First hints for such a scenario have been shown for the 
Wolf-Rayet stars WR\,124 with its nebula M1-67 and the LBV 
He\,3-519 \citep{2015wrs..conf..167W}.  
The M1-67 WR nebula is one of few if not the only one that has a 
bipolar morphology and a size of only 2pc. As~described above 
these are very typical values for LBV nebula. One might picture 
WR 124 as an old LBV that has just left the LBV and entered the WR phase, 
matching well to its current WN 8 spectrum. The~scenario 
for He\,3-519 might be just 
reversed, the~stars is an LBV that is turning into an WR right now. This would
also explain why no S Dor variability is seen for that star.
Its current spectra type is already that of a WN 11. The~ 
nebula is only weakly bipolar and with 2 $\times$ 2.5 pc rather large for 
an galactic LBV see HST image in Figure~\ref{fig:neb}. 
It looks more like an old LBV nebula that by inflation via 
the strong WR stars wind has already increased its size. Doing so also caused 
bipolarity to fade o\citep{2015wrs..conf..167W}.   
For a more general review about WR stars see the 
contribution by Kathryn Neugent and Philip Massey in this~volume.

\section{Links of SN Impostors and~LBVs}
In recent years several projects and monitoring surveys that search for 
supernovae found what has become known as {\it SN impostors}. 
These transients show spectra similar to core-collapse SN, especially~of the
type SNIIn, but~are generally significant fainter than core-collapse SN.
SN impostors show lightcurves quite different from all core collapse SN,
sometimes even showing strong fluctuations on short timescales some time
after the initial eruption, see e.g.\ \cite{Pastorello2010}.
It is interesting to note, that the brightest impostors events even overlap
in energy with the faint SN IIP, e.g.\ \cite{Arnett1989, Pastorello2007_2}.
A very tempting and likely explanation is to identify at least a subset of
these SN impostors with giant eruptions or even S\,Dor variabilities of LBVs~\cite{2005AA...429L..13W} in distant galaxies. 
Note in this context that while a LBV giant eruption will look like a 
SN impostor, not all SN impostors might indeed be LBV giant eruptions!

With the current list of about 40 SN impostors~\cite{Bomans2019},
light curves and spectra during the eruption, and~especially the pre- and
post-eruption behavior imply at least two different object classes are
summarized in the name Impostor:  
the transients with strong narrow emission lines and erratic lightcurves
with secondary, smaller outbursts following the first eruption, and~the  
transients, which~are followed with less than a decade by a true 
supernova explosion, e.g.~\cite{Pastorello2019}.   This diversity of the
lightcurves and spectra of the transients denoted as SN impostor was also
noted by Smith~\cite{Smith2011_5}.  
It is potentially important, that the rise of the eruptions can be very steep  
\citep{Pastorello2010, Bomans2016, Bomans2019}, putting interesting limits 
on the kinetic energy and size scales~evolved.  

SN impostors will be discussed in detail in this volume by Kris~Davidson.

\section{Multiplicity of~LBVs}

{ In recent years it became clear, that a significant part of massive stars are
born in double (or multiple) systems.  A~detailed analysis of the results
from the FLAMES-Tarantula survey lead Sana~et~al.\ \citep{Sana2012} to
the following percentages for massive stars: effective single (real single
stars or wide binaries without significant interaction $\sim$29$\%$,
stellar merger $\sim$24$\%$, accretion and spin up or common envelope evolution
$\sim$14$\%$, and~envelope stripping $\sim$33$\%$ \citep{Sana2012}. Therefore,
about $\sim$71$\%$ are affected by binary interaction. Alternatively, if~one sees
the result of mergers as apparent single stars for most of their lifetime,
then only $\sim$47$\%$ of the massive stars should show a companion. 
The same data also imply that the numbers of equal mass binaries are lower than 
unequal mass pairs. The~ratio goes up to $\sim$50$\%$ at M$_2/$M$_1 = 0.3$, the~
lowest mass ratio probed by their data \citep{Sana2012}.

This immediately implies that a sizable number of LBVs should have binary
companions. The~idea, that binary star evolution is linked to the LBV
phenomenon is quite old, see e.g.\ Gallagher (1989) \citep{Gallagher1989}.
More recently,  the~idea of mergers triggering giant eruptions (or
being one path to SN impostors) gained some interest,
e.g.\ \citep{Justham2014}.
Still, the~observation of binary companions of LBVs are
difficult, due to the large luminosity of the primary, and~its strong
stellar wind, which~both limits spectroscopic searches. Direct imaging
searches only cover relatively large separations, and~only few LBVs
are analyzed with stellar interferometers, yet. A~search for X-ray only covers
situations in which colliding winds can occur, and~may be in part contaminated
by the X-ray emission of circumstellar~nebula.

The current state on observed stellar companions to LBVs is the following: 
As shown in chapter \ref{eta} $\eta$ Carinae show strong signs of being 
a binary star with a massive, hot companion stars. 
HD 5980 was first reported as an exlipsing LBV Wolf-Rayet binary system that  
showed an LBV like eruption \citep{1994IAUC.6099....1B}. 
Koenigsberger~et~al. \citep{2014AJ....148...62K} report new analysis which is
consistent with the system being more complex and multiple: a double binary
scenario and manifests a quadruple system. 
The LBV candidate [KMN95] Star A (= 1806-20) showes double He lines \citep{Figer2004} 
but single emission lines, implying a dense stellar wind for the primary, similar to
the case of $\eta$ Carinae. 
MCW 314 shows clear indications in its lightcurve and its radial velocity curve for
having a lower luminosity supergiant companion \citep{Lobel2013}.
If the wide companion candidate \citep{Martayan2016} is truly bound, than~
MWC 314 would be a hierarchical triple star. 
The LBV HR Car was observed with stellar interferometry and strong
indications of a companion was found \citep{Boffin2016}. The~companion star
appears to be relatively low mass (below $\sim$15M$_{\odot}$). 

A search for wide companions based on natural seeing, AO assisted imaging,
and archival HST imaging of 7 galactic LBVs, LBV candidates, and~some related
objects yielded one star with potential companion (MWC 314) and no
apparent bound companions for the 5 other LBVs and LBV candidates the Pistol star, 
HD 168625,\,HD 168607,\,MWC 930,\,and [KMN95]\,Star A (=1806-20) \citep{Martayan2016}. 
PSF subtracted HST images used in the study of LBV nebulae in the LMC
by Weis \citep{2003AA...408..205W} also showed no apparent companion stars, 
but only relatively large projected orbital distances could be probed ($>$0.1\,pc).

A X-ray archival survey (using XMM-Newton and CHANDRA X-ray satellites)
of 31 LBVs, LBV candidates, and~related objects
was performed by Naze~et~al.\ \citep{Naze2012}.  X-ray emission may indicated
colliding winds in a binary, but~(softer) X-ray could also be created in a
circumstellar nebula, see e.g.\ Weis \citep{2004AA...415..595W} for the case of
$\eta$ Carinae.  The~survey of Naze~et~al. yielded 4 detection ($\eta$
Carinae, W243 (= Westerlund 1 \#243), MSX6C G026.4700+00.0207
(= GAL 026.47+00.02), and~Schulte \#12 (= Cyg OB2 \#12).
Two more are labeled doubtful candidates (GCIRS 34W, and~ GCIRS 33SE) by the authors. 
This result also implies a long list of 25 non-detections, which
includes confirmed LBVs like P\,Cygni, the~Pistol star, and~FMM 362.  While acknowledging
their rather heterogenous data base, the~authors suggest that their
detection rate is consistent with a binary fraction between 26$\%$ and 69$\%$,
roughly consistent with that of other classes of hot, massive~stars. 

Given the very different methods used, and~the therefore very different
orbital radii and mass (and luminosity) ratios probed up to now, it is hard
to derive a reliable result on the binary fraction for LBVs as a class.  
An additional problem are the very different LBV input lists used in the different 
searches. There are clearly several good cases for binary companions of LBV stars.
Still, we regard the actual binary fraction of LBVs as currently very uncertain, but~most 
likely around $\sim$20$\%$ for the confirmed LBVs. This would be somewhat lower 
than the binary fraction for other classes of massive stars like O supergiants 
or Wolf-Rayet stars.  If~this estimate of the binary fraction is correct, 
it may hold important clues for the evolutionary pathways leading to LBVs.
}
 
 
\section{LBV and Their~Neighborhood}

Smith \& Tombleson\citep{2015MNRAS.447..598S} analyzed the location of 
LBVs in comparison to their surrounding and concluded that LBVs in MW and 
LMC are isolated, and~
not spatially associated with young O-type stars. This would imply 
a complete change of the standard view of the evolution of LBVs, 
clearly a far reaching claim, which needed further investigation.
Humphreys~et~al.\ \citep{2016ApJ...825...64H} analyzed the location of a 
sample of LBVs in M\,31,
M\,33, and~the LMC in comparison too other massive main sequence and 
supergiant stars. 
With this large and more coherently selected sample,Humphreys~et~al.
\citep{2016ApJ...825...64H} concluded that LBVs are associated with supergiant
stars and are neither isolated or preferentially run-away stars.  
Separating the more massive classical and the less luminous LBVs, 
the classical LBVs have a distribution similar to the late O-type stars, 
while the less luminous LBVs have a distribution like the red
supergiants.
Smith  \citep{2016MNRAS.461.3353S} questioned the results of this analysis 
and reiterated the results of his analysis.
Davidson~et~al.\ \citep{2016arXiv160802007D} shortly after showed that the 
statistical analysis methods use in \citep{2016MNRAS.461.3353S} are flawed.  
Independently, Aadland~et~al. \citep{2018AJ....156..294A} performed a very
similar analysis and came to similar  conclusions as Humphreys~et~al. 
\citep{2016ApJ...825...64H}, that the stellar environment of LBVs is the 
same as for supergiants.
It is still be worth noting, that  the Aadland~et~al. sample is not a clean
LBV sample, but~contains many B[e] supergiants. Note that
this point was also pointed out by Kraus in her review paper on B[e] in this
volume.  In~a recent paper Smith
\citep{2019MNRAS.489.4378S} gravitated to the interpretation by Humphreys~et~al.\ of LBV locations within (or near) their birth association.
Just lately with an analysis of GAIA data~\cite{Ward2019}, strong evidence was
presented, that OB stars form not preferentially in bound clusters, but~in a
continuous distribution of gas densities, at~many locations of the birth cloud.
This~view is also supported by recent simulations which also favor a
hierarchical formation model for the formation of OB stars as a result of
the fractal structure of the birth clouds, contrary to a monolithic collapse.
In this picture many different stellar neighborhoods of massive stars would be
natural, also consistent with our~results.

\section{The Population of~LBVs}

As mention above the first reports on Var\,2 in M\,33 was already in the 
1920ties 
marks the first identification of an LBV-at that time without the
knowledge that it is and what LBV are. 
Since the Studies by Hubble \& Sandage 1953 \citep{1953ApJ...118..353H} and 
Sandage \& Tamman 1974\citep{1974ApJ...194..559S} we know that both M\,31 and 
M\,33, as~well as NGC\,2403 are known to host several LBV and LBV candidates. 
S\,Dor added the LMC into the list of LBV host galaxies.  The~LMC 
has a remarkable population of LBVs \citep{1979ApJ...232..409H,
1988BICDS..35..145L}.
Bernhard Wolf and his group in Heidelberg studied 
various LBVs and LBV candidates in several galaxies and with this 
first larger sample was able to identify the above mentioned amplitude 
luminosity relation. Beside that they found several LMC LBV candidates and   
confirmed many LBVs by observing their S\,Dor Cycles  like  R\,127
\citep{1988IAUS..132..557S}, R\,110 \citep{1989AGAb....3..115S}. They also
noticed an inverse P\,Cygni profile in the spectrum of S\,Dor \citep{1990AA...235..340W}.
and added HD\,160529 to the galactic LBVs. Last but not least the group
also identified with R\,40 the very first LBV in the 
SMC \citep{1985AAS...61..237S,1993AA...280..508S} . 
Other LBV host galaxies now known are locally IC\,10 and further out 
are the M\,1 group members M\,1, NGC\,2366. LBV and LBV candidates
are reported also in M\,101, NGC\,300, NGC\,247, NGC\,6822, NGC\,4414, and~
IC\,1613, just to name the most important~galaxies.  

In Table~\ref{tab:lbvlist} a list of known LBV and LBV candidates is given. 
True LBVs are those stars where the membership is clear since a complete S\,Dor
cycle has been observed, this is not the case for the LBV candidates. 
For LBVs that had a giant eruption, those are classified separately 
and named giant eruption LBVs (or $\eta$\,Car Variables) to distinct them 
from LBVs with S\,Dor variability~only.

Several more stars have for the one or other reason be classified as LBVs
by one or more authors, but~show no clear hints like S \,Dor cycle or giant
eruption.  
For the Milky Way the objects HD\,80077 and Schulte\,12 the new GAIA  
parallax moves both to a closer distance and to a lower luminosity. {Still, 
the~GAIA parallaxes are at this time (GAIA DR2) prone to several systematics 
\citep{Lindegren2018_2, Lindegren2018}.

\begin{table}[H]
\caption[]{LBVs and LBV candidates in alphabetic order. 
Giant eruption LBVs are italic. Objects marked bold have an 
(optical) emission LBV nebula.  
Except for the Milky\,Way and LMC which have a to large number of 
objects, references are given.}
\label{tab:lbvlist}
\begin{center}
\begin{tabular}{|l|l|l|l|}
\hline 
\textbf{Galaxy} & \textbf{LBVs} &  \textbf{LBV Candidates} & \textbf{References} \\
\hline
Milky\,Way  & {\bf\,AG\,Car}, \textbf{ \emph {$\eta$\,Car}}, FMM\,362, & 
BD+143887, BD-13\,5061, B[B61]\,2,  & \\
{} &  [GKF2010]\,MN44,   &  G025.520+0.216, G79.29+0.46, GCIRS\,16C,    & \\
{} & [GKM2012]\,WS1,      & 
  GCIRS\,16NE, GCIRS\,16NW, GCIRS\,33SE,    & \\        
{} & HD\,\,168607, HD\,160529,        & GCIRS\,16SW,  [GKF2010] MN58,    & \\ 
{} &  HD\,193237, {\bf  HR\,Car} , & [GKF2010]\,MN61, [GKF2010] MN76,  &  \\           
{} & 
{\bf LBV G0.120-0.048}, MWC\,930,    &  [GKF2010]\,MN 80, [GKF2010]\,MN83,     & \\
{} & \textbf{ \emph {P\,Cygni}}, V481 Sct,  & [GKF2010] MN96, [GKF2010] MN112,    & \\
{} & W243, {\bf WRA\,751} & [GKF2010] WS2, {\bf HD\,168625}, HD\,316285,     &  \\        
{} & & HD\,326823, {\bf He\,3-519}, IRAS16278-4808,  & \\   
{} & &  IRAS19040+0817, J17082913-3925076,  & \\
{} & & [KMN95] Star A, & \\
{} & & MSX6C G026.4700+00.0207, {\bf Pistol star}, & \\
{} & & {\bf Sher\,25}, WR\,102ka, & \\
{} & &   WRAY 16-137, WRAY 16-232   & \\ 
\hline
LMC  & HD\,269216,R\,71, R\,85,  & HDE\,269582, {\bf S\,61}, {\bf
  S\,119}, {\bf Sk-69$^{\rm o}$ 279} & \\
{}   & {\bf R\,127}, {\bf R\,143}, R\,110, S\,\,Dor,&      & \\       
\hline
SMC   & R\,40 & (R\,4), (R\,50)& \citep{1993LNP...416..280S} \\
\hline  
M\,31   & AE\,And, AF\,And,  &  J003910.85+403622.4, J00441132+4132568, &
\citep{2015PASP..127..347H,Humphreys2017} \\ 
& LAMOSTJ0037+4016,  & M\,31-004425.18, M31-004051.59   & \citep{2019ApJ...884L...7H} \\
{} &  UCAC4\,660-00311,  &    & \\
{} &  Var\,A-1, Var\,15    &   &     \\ 
\hline    
M\,33   &  Var\,B, Var\,C, Var\,2, Var\,83, & GR 290, 
[HS80] B48, [HS80] B416 ,  & \citep{Humphreys2017}  \\ 
{} & & [HS80] B517, J013228.99+302819.3,  & \\
{} & &  J013235.21+303017.4,J013317.01 + 305329.87, & \\
{} & &  J013317.22+303201.6, J013334.11+304744.6,     & \\
{} & &  J013337.31336+303328.8, J013351.46+304057.0, & \\
{} & &  J013354.85+303222.8, J01342475+3033061,      & \\
{} & &  J01342718+3045599, J013432.76+304717.2,    & \\
{} & &  J013459.36+304201.0, J01350971+3041565,     & \\
{} & &   M33C-5916, M33C-10788, M33C-15235,  & \\ 
{} & &  M33C-16364, M33C-21386, UIT 008       & \\
\hline  
NGC\,2403   & {\it SN\,1954J=V12},    & V\,22, V\,35, V\,38 &
\citep{1968ApJ...151..825T} \\  
& SN\,2002kg=V37 & & \citep{Humphreys2019, 2017ApJ...848...86H} \\  
\hline  
NGC\,1058   & {\it SN\,1961V} &  & \citep{1964ApJ...139..514Z,1989ApJ...342..908G} \\  
\hline  
NGC\,2366   & NGC 2363\,V1 & & \citep{2001ApJ...546..484D} \\  
\hline  
M\,101 &  & J140220.98+542004.38, V\,1, V\,2, V\,4, V\,9, V\,10  &
\citep{1983AJ.....88.1569S, 2015AJ....149..152G} \\
\hline  
M\,81 &  & I\,1, I\,2 & \citep{1984AJ.....89..621S, Humphreys2019}\\
\hline  
IC\,10 & & unnamed  & \citep{2003IAUS..212..160C}\\
\hline  
NGC\,300  & & B 16 & \citep{2002ApJ...577L.107B}\\
\hline  
NGC\,6822 & & unnamed & \citep{1980ApJ...238...65H} \\ 
\hline  
NGC\,4414  & & unnamed & \citep{1998ApJ...505..207T} \\
\hline  
IC\,1613 & &  V\,39, V1835, V2384, V3072,     &
\citep{1980ApJ...238...65H} \\
& & V3120, V0416, V0530,  & \citep{2000AA...363...29A,2010AA...513A..70H} \\
\hline  
UGC\,5340 & & unnamed & \citep{Pustilnik2008, Bomans2011, Pustilnik2017} \\
\hline  
NGC\,3109 & & unnamed & \citep{Bomans2012} \\ 
\hline
\end{tabular}
\end{center}
\end{table}

\noindent

Therefore the distance of at least the Schulte\,12 is still not settled yet 
\citep{2019MNRAS.484.1838B}.}
According to Humphreys~et~al. \citep{2016ApJ...825...64H} in the LMC  R\,66,
R\,74, R\,123  are B[e], R\,149 is an Of star, HD\,269604 an  A supergiant and 
HD\,34664 as well as  HD\,\,38489 are B[e]sg. Neither are R\,81, R\,84, R\,99,
R\,126 LBVs. The~SMC object  R\,50 is a B[e]sg while R\,4 is a spectroscopic
binary system with one B[e]sg. 
Finally HD\,5980's activity is more like a giant eruption but this most likely
due to a binary interaction, see chapter on Multiplicity of LBVs below. 
Therefore 
its seen as a giant eruption LBV candidate with the above caveat. 
Just recently Humphreys \citep{Humphreys2019}  report that the following 
objects are not LBVs: I\,8 in M\,81 is an F supergiant, furthermore V\,52 in 
NGC\,2403 and I\,3 in M\,81 are foreground objects and not even part of those galaxies!
HD\,168625, He\,3-519, Pistol star, and~Sher\,25 are LBV candidates due  
to the fact that they posses a circumstellar nebula. They however might 
indeed be LBVs, as~members of what van Genderen classified as a group 
ex/dormant LBVs, just currently not showing any variability. 
Besides the variability searches, there is also a consistent search of
luminous  emission line stars using two or more broad band colors, H$\alpha$
as detection and [OIII] as veto filter,  done as part of the NOAO Local Group
Survey \citep{Massey2007,  Massey2007b}  and independently by our group
\citep{Burggraf2007}. Both~searches covered M\,31, M\,33, NGC\,6822, IC\,10,
Wolf-Lundmark-Melotte, Sextans\,A, and~Sextans\,B and finding very few
candidates in the dwarf~galaxies. We checked also NGC\,3109, a~low
metallicity galaxy forming a subgroup with Sextans\,A and Sextans\,B at the
fringes of the Local Group. We found only one candidate \citep{Bomans2012},
similar to the low candidate  numbers for the low metallicity dwarfs in the
Local Group.An~earlier attempt with the same idea to detect very luminous stars,
which are  strong H$\alpha$ line emitters (either from strong mass loss, or~
from a circumstellar  nebula), which are faint or absent in [OIII] (no stellar
emission line, and~faint for  circumstellar nebulae of CNO 
processed material)
was done by the Heidelberg group  \citep{Spiller1992} for M\,33, M\,81,
NGC\,2403, and~M\,101, but~was not published.
We used e.g.\ these data to
complement our list of good candidates for spectroscopy in M\,33
\citep{Humphreys2014, Humphreys2017}.
It is interesting to note here, that coordinated searches for variable stars
(in particular not only analyzing the Cepheids) is done only for small number of
massive local galaxies since the photographic  plate area.  A~new effort is
ongoing with the LBT and yielded already interesting results
\citep{Humphreys2019}. Our group is currently working on a search for LBV
and related objects in several nearby galaxies.

\section{LBVs in Low Metallicity~Systems}

The situation is even worse for LBVs in lower mass galaxies.  Detections are rare 
as metal-poor also implies low mass and even in actively starforming dwarf galaxies 
the numbers of massive stars are more limited as in large, massive spirals. The~
SMC, for~example, on~has only one confirmed LBV, see Table~\ref{tab:lbvlist}. 
An interesting LBV candidate is V\,39 in the low metallicity Local Group dwarf 
galaxy IC\,1613. Detailed analysis of its spectrum shows some patterns 
similar to other LBVs, but~is also consistent with that of a sgB[e] 
star \citep{2010AA...513A..70H}.


Besides the
aforementioned  Local Group galaxies, there are only chance detections up to
now, including the exceptional case of NGC\,2366\,V1.  NGC\,2366\,V1
\citep{Drissen1997, Drissen2001, Petit2006} is located in a dwarf galaxy with
a metallicity below 1/10 solar.  Its ``outburst'', with~a change of only
$\sim3$ mag \citep{Drissen1997}  was probably not really a giant eruption).
Neither did it follow the classical S\,Dor pattern since it turned bluer (not redder) with increasing brightness. 
 
The transient in UGC\,5340 (DDO\,68) was again a chance detection
\citep{Pustilnik2008}. The~brightening  is 1 mag, and~here again a blueing
during the bright state is
visible \citep{Bomans2011,Pustilnik2017}.  The~galaxy  is a morphologically
peculiar, low mass system, which has with $\sim1/30$ solar (log (O/H) = 7.12)
one of the lowest gas-phase metalicities in the nearby universe (distance
12.6 Mpc \citep{Sacchi2016}. The transient in PHL\,293B (= SDSS\,J223036.79-000636.9) \citep{Izotov2009} 
{is difficult to study mainly due to its distance of $\sim$25 Mpc.} The host
galaxy is a dwarf galaxy and more metal-poor than the SMC (log O/H = 7.72). 
The transient discovery spectrum shows clear P\,Cygni profiles, but~no details on 
the temporal variability were known, only 2 spectra (one without and one with P\,Cygni 
profiles).

\begin{figure}[H] \centering
\includegraphics[width=13 cm]{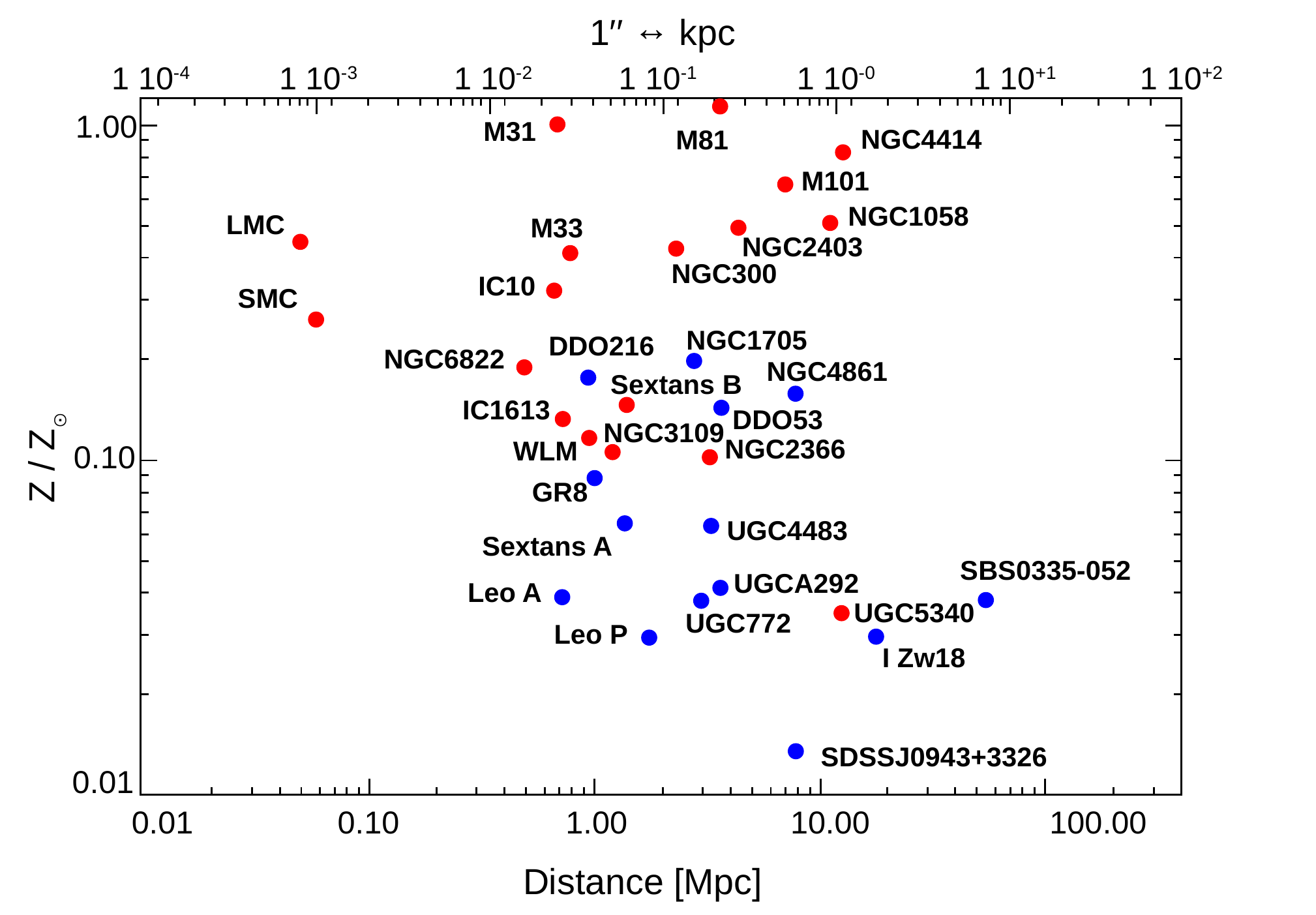}
\caption{Plot of the gas-phase metallicity of nearby galaxies versus their
distance and spatial resolution.   Only a selection of the  galaxies in the
Local volume are plotted, but~the sample is complete for the significantly
starforming galaxies  in the Local Group.  Metalicities of the inner disk are
chosen for the spiral galaxies with metallicity gradients,  the~metalicities
of stars in the outer disk of these galaxies can be a a few faction of tens
solar lower.  Galaxies with LBVs and/or LBV candidates are plotted as red
dots, the~other galaxies are plotted as  blue dots. Plot was adapted and
updated from \citep{Bomans2011}.}
\label{fig:Zdplot}
\end{figure}   

 { An additional spectrum brought the time baseline to 8 years and proved 
temporal variations of the broad stellar lines \citep{Izotov2011}.
While being an interesting object, which may acquire LBV candidate status 
with a longer term photometric and spectroscopic monitoring, but~the currently limited data } 
makes the label LBV for this object a bit premature.  
As similar problem is the transient in the galaxy SDSS\,J094332.35+332657.6 \citep{Filho2018}, 
an~apparent stellar transient in an very low mass and extremely low metallicity (log (O/H) =
7.03) galaxy at a distance of $\sim 8$ Mpc. Only a very limited historical record is 
available, and~therefore the LBV nature of the transient is quite unclear.  
It may be interesting to note here that the LBV GR\,290 (=~Romano's star) in M\,33 also
shows  spectra variability, but~not consistent with an S\,Dor pattern, see
Maryeva~et~al., this volume.  The~star is located in the outer regions of the disk of
M\,33 (r= 4.3 kpc from the center of  M\,33.  The~observed metallicity gradient
\citep{Rosolowsky2008} therefore implies a low metallicity  of log (O/H) =
8.2 (roughly between LMC and SMC \citep{ToribioSanCipriano2017}) for the
star. Note in that context  that metallicity gradients are a common feature in
spiral galaxies, e.g.\  \citep{Sanchez2014, Bresolin2019}, so large spiral
galaxies do not have one fixed metallicity.

Another intriguing object was detected by as a point source with high
velocity dispersion in H$\alpha$ Fabry-Perot observations of the local
(D$\sim$2.6 Mpc), low metallicity dwarf galaxy UGC\,8508 \citep{Moiseev2012}.
An~intermediate dispersion spectrum of the source shows a bright H$\alpha$
line with broad wings, a~relatively strong Fe\,II $\lambda 4924$ line, but~
also a strong He\,II $\lambda 4686$ line.
The classification of the authors as a massive star with strong mass loss is
convincing, but~if it is indeed a good LBV candidate is more uncertain,
given the high temperature (and/or hard radiation field) implied by the
presence of the strong, narrow He\,II line.

We detected another unusual point source \citep{Kleemann2019} in NGC\,1705,
a starburst dwarf galaxy at D$\sim 5$ Mpc  with a metallicity similar to
the LMC. The~spectrum shows several very strong (and split) forbidden 
emission lines, all showing an expansion velocity of 50 km s$^{-1}$, and~an
underlying spectrum of the source is that of an A supergiant.
Again this is a massive star with an expanding circumstellar
bubble, but~its exact nature is not determined yet. 

The starter for the question, how many galaxies do we know in the local
universe  (e.g. the Local Volume = D < 11 Mpc) based on the classic compilation of
\citep{Tully2009}. There is an obvious  distance limit  when using photometry
from the ground, especially historic photographic plate material for  long
term light curves to identify LBV candidates.  This limit is depending on
seeing, size of  the telescope used, and~the detector.  The~limits for
photographic plate work is about 7 Mpc  (the distance of M\,101)
\citep{1968ApJ...151..825T}, and~is for most telescopes more like $\sim4$ Mpc
(the  M\,81 and IC\,342 groups in the north, and~Sculptor and Fornax groups in the
south).  Obviously with CCDs and good seeing this can be extended (and/or the
quality of the photometry  improved), but~access of older CCD data is tricky,
if the observatory does not run a well maintained  archive.  Clearly, HST and
in the near future EUCLID and JWST, can go much farther out, but~it  gets hard
beyond 20 Mpc (especially due to the crowding of stars). 

Low metallicity LBVs are especially interesting, since the
metallicity can influence { opacity in the interior of the stars and in the 
wind}. Metallicity also affects the path of the evolutionary tracks
(at which mass stars still go RSG, return to the blue, or~
go through a LBV phase with  S\,Dor-like variability and  instability that 
caused these, etc...) Furthermore rotation rate, binary fraction, and~
potentially IMF as well as magnetic fields are important.

Several of this markers of LBV candidates are directly, or~indirectly
influenced by metallicity.  Mass loss e.g.\  \citep{Tramper2011, Tramper2014,
Garcia2019}, emission lines of heavy elements (e.g. photospheric or wind
emission lines  of FeII, FeIII, [FeII], then HeI, and~[NII] in a circumstellar
nebula) \citep{Humphreys2014, Humphreys2017}, and~variability due to  the
metallicity dependence of the instabilities involved (see above).  It could be
that at low enough metallicity,  massive stars behave differently, e.g.,~not
showing an typical S\,Dor  variability pattern anymore.  The~cases of V1 in 
NGC\,2366 \citep{Drissen1997} and the transient in  UGC\,5340 (DDO\,68) 
\citep{Bomans2011,Pustilnik2017} hints towards and seem to support such a
scenario. 
No coordinated search for luminous variable sources in a sample of low
metallicity dwarf galaxies  outside the Local Group was done yet.  A~pilot
search on a few selected very low metallicity galaxies  was reported by
\citep{Bomans2011b} using HST archival data. While~there are several
interesting candidates  of luminous stars with signs of variability and in some
cases H$\alpha$ emission, the~data yield not enough proofs to claim LBV
candidates.  Figure~\ref{fig:Zdplot} demonstrates one of the problem, very low
metallicity galaxies are rare  and spatial resolution poses severe problems
for ground based studies beyond $\sim$5 Mpc,  requiring HST time.  This
aspect may improve with the upcoming EUCLID mission and more in the future by
WFIRST, and~is alleviated somewhat by the  improving image quality of the
large survey instruments, link e.g.\ SUBARU SuprimeCAM, DECam, and~hopefully LSST.
Another problem is the metallicity-luminosity relation, which implies that
low metallicity is in the local universe  the exclusive regime of dwarf
galaxies.  Therefore, even in a burst of starformation the absolute number
of massive  stars produced is, during~a short time frame only, still
comparable to the production rate of a massive spiral galaxy.  With~the
current data situation it is to early to speculated on trends of LBV numbers
and LBV nature at low metalicities,  but~as noted above, it is intriguing to
see so many LBVs and LBV candidates in the LMC. With~at the same time nearly non
in the SMC.

\section{Summary and~Conclusions}

The Luminous Blue Variable phase is a short phase in the life of massive
stars. It may be passed by stars with an initial mass as low as
21\,M$_{\odot}$. LBVs have a specific variability the S\,Dor variable, can~
undergo giant eruption and have very high mass loss rate. The~one and only way 
to pinpoint and truly classify LBVs is by the variability and/or giant
eruption This asks for the detection of at least one S\,Dor cycle  
the star passes or to catch it in a giant eruption. These variabilities 
also subdivide the  LBV class in classical (S\,Dor variable) LBVs and giant
eruption LBVs.
With the variability as the only clear classification method many
LBVs in a quiescence state might be overlooked and not be identified as such.
It is therefore not trivial to describe the LBV population in a
galaxy. In~that connection not knowing the true amount of LBVs and non-LBVs
makes it hard to give an estimate for the real duration of the LBV phase. This
again is directly linked to uncertainties of the total mass loss rate of 
massive stars. 
Even small changes of the phase length are linked to large changes in the mass
total loss of the stars, given LBVs have very high mass loss rates. Last but
not least that implies that the final mass of stars that pass a LBV 
phase could be much lower as thought so far. In~that case this would even 
effect amounts and ratios of different SN types. 

The path is therefore clear, to~better characterize the LBV population and
the underlying physics more long-term variability studies of nearby
galaxies are needed.  Spanning the parameter space especially towards lower
metalicities will potentially clarify the importance of {opacity effects and 
rotation} for the S\,Dor variability. Also analyses of the long-term variability 
of massive stars in all the most metal-rich spiral galaxies in the Local Universe 
are not really done yet.  First attempts are already ongoing, partly using data 
from well maintained archives, and~the time-domain section of future
large survey projects like LSST will be a major step forward. This will
also be true for a better understanding of eruption LBVs.  Another promising
avenue will be the ``archaeology'' of the mass-loss of LBVs and related stars
using their circumstellar nebulae.  In~this way, information on energy, mass,
and chemical composition of earlier mass-loss of the stars can be
investigated, again providing clues about the underlying mechanism of
instability and the evolutionary state of the stars.  With~the rise of
integral field spectrographs, even with AO support (e.g.\ MUSE at the
ESO/VLT), such analyses should be possible in all Local Group galaxies 
{ and the nearest galaxy groups. First such analyses are already 
appearing for galaxies in the Scultor group: NGC 300 \citep{Roth2018} and 
NGC 7793 \citep{Wofford2020}.}
An unfortunate weakness in the currently available instrumentation are
high-dispersion spectrographs fed by long-slits and IFUs, an~important
capability for kinematics/energetics of nebulae, which is becoming
rare \citep{Bomans2014} at the intermediate and large telescopes.
High-multiplex spectroscopic survey instruments at large telescopes, like
e.g.\ Hectospec at the MMT, and~soon MOONS and 4MOST at ESO telescopes, 
as~well as WEAVE at the WHT, can be very useful tools to set LBVs in context 
to their massive star environment, 
as they are capable of providing good quality spectra for many photometrically 
selected LBV candidates (as well as other supergiants). This still requires 
that starforming, nearby galaxies 
will be targeted in the upcoming large surveys at these facilities.
Taking this all together, one can be optimistic, that in the coming years
many more good quality observational data will be available to improve our
understanding of the LBV phenomenon and its importance for the evolution
of massive stars.





\vspace{6pt} 



\authorcontributions{KW planned the review structure. Both authors contributed
  then roughly equally to writing this review, with slightly different relative amounts depending on the topics covered in each chapter.
} 

\funding{This work was supported through the Astronomical Institute of the
    Ruhr University Bochum, the Ruhr University Research Department Plasmas with Complex Interactions, and DFG Research Unit FOR 1254.
}

\conflictsofinterest{The authors declare no conflict of interest.
 } 


\acknowledgments{The authors thank Roberta Humphreys for many discussions,
and her many comments and suggestions for this text; Kris Davidson for
several helpful comments; Jochen Heidt and Alexander Becker for their
contributions to the LBT AO observations of P\,Cygni; and our students for
their contributions to many aspects of LBV related research here at Bochum
during the last years. Thanks also got to two anonymous referees whose suggestions improved this paper.}



\newpage
\abbreviations{The following abbreviations are used in this manuscript:\\

\noindent 
\begin{tabular}{@{}ll}
BSG & Blue Supergiant \\
ESO & European Southern Observatory\\
HRD & Hertzsprung-Russell Diagram\\
HST & Hubble Space Telescope \\
JWST & James Webb Space Telescope \\
LBV & Luminous Blue Variable\\
LMC & Large Magellanic Cloud \\
LSST & Large Synoptic Survey Telescope \\
MMT & (converted) Multi Mirror Telescope\\
NOAO & National Optical Astronomy Observatory \\
RSG & Red Supergiant \\
SMC & Small Magellanic Cloud \\
SN & supernova \\
WFIRST & Wide Field Infrared Survey Telescope \\
WHT & William Herschel Telescope\\
WR  & Wolf-Rayet star 
\end{tabular}}

\externalbibliography{yes}
\bibliography{kweis,djb}




\end{document}